# THE LIMIT $m_q \to 0$ OF PERTURBATIVE QCD

F.V. TKACHOV

*Institute for Nuclear Research of the USSR Academy of Sciences, 60th October Prospect, 7a, 117312 Moscow, USSR*

The effect of light quark masses, as seen from short distances, is discussed. The large logs of $m_q/\Lambda$ are shown to simulate "vacuum condensates".

*1.* The aim of this Letter is to show that the role of light quark masses has heretofore been strongly underrated. Indeed, according to the popular philosophy the effect of light quark masses should be relatively unimportant, while the leading role is ascribed to the complex non-perturbative vacuum which is considered as only lightly affected by $m_q \neq 0$ [1]. Nobody, however, has as yet seriously studied the effect of large logs of $m_q/\mu$, $\mu$ being the renormalization parameter, which appear in perturbative calculations. Below we argue that these logs, when resummed, can drastically change the naïve understanding of the limit $m_q \to 0$, so that the latter should rather be considered in the spirit of Bogoliubov's quasi-averages [2]. By this we mean that the real QCD near $m_q \approx 0$ may differ qualitatively from the theory with massless quarks if such a theory exists (which is far from being obvious).

To study the limit $m_q \to 0$ of the full theory one must *first* expand the full Green functions in $m_q$ and *only then* try, if one wishes, to calculate the resulting coefficients by expanding them in $\alpha_S$. For obvious reasons such a procedure is hardly possible. Instead one usually proceeds in a completely opposite direction: one takes the limit $m_q \to 0$, graph-wise, *after* expanding in $\alpha_S$, so that $\log(m_q/\mu)$'s are always suppressed by powers of $m_q$.

It would be preferable to try to see the whole truth by resumming, as best we could, those logs prior to making conclusions. Needless so say, this is no substitute for a non-perturbative analysis, but it is reasonable to expect that this way corresponds better to the original intention.

*2.* The Bogoliubov-inspired [3] extension principle (EP) for studying asymptotics of Feynman integrals [4] in conjunction with the $R^*$-operation for removing a class of infrared divergences in the presence of ultraviolet ones [5], offer a means to study the structure of the $m_q \to 0$ expansion of Feynman integrals straightforwardly and to all orders of perturbation theory (PT).

To avoid irrelevant combinatorial complications we take a very simple example, but which retains all the essential features of the full proof[1]. Our example is similar to the one considered in [4] and our reasoning closely follows [6]. Therefore only formal steps of the derivation will be indicated below. The MS-scheme [7] is used throughout the paper.

Consider a free theory with a massless field $\varphi$ and a massive one $\Phi$ in 4 dimensions, and the current $j(x) = :\varphi(x)\Phi(x):$. Consider

$$\Pi(Q^2, m^2) = i\int e^{iqx} d^D x \langle RTj(x)j(0) \rangle_0$$
$$= \int \frac{d^D p}{(2\pi)^D} \frac{1}{(p-Q)^2(m^2+p^2)} - C.T. \quad (1)$$

where $Q^2 = -q^2$, and the rotation to Euclidean region has been performed on the r.h.s. of (1). The UV counterterm in (1) *does not depend on $m$*. Formal expansion gives:

$$\text{Eq.}(1) = \int \frac{d^D p}{(2\pi)^D} \frac{1}{(p-Q)^2} \left[\frac{1}{p^2} - \frac{m^2}{p^4}\right] - C.T. + o(m^2). \quad (2)$$

The UV counterterm still subtracts the UV divergences from the first term in (2). The second term has no UV divergences but a logarithmic IR one. Following EP [4], to obtain a correct[2] expansion in $m$, we must add a counterterm into (2):

$$-\frac{m^2}{p^4} \to -\frac{m^2}{p^4} + z(m)(2\pi)^D \delta(p). \quad (3)$$

(**NB** Derivatives of $\delta(p)$ need not and should not be added [4], [6].) To fix the unknown $z(m)$ we demand that

$$0 = \int \frac{d^D p}{(2\pi)^D} \psi(p) \left[\frac{1}{m^2+p^2} - \frac{1}{p^2} + \frac{m^2}{p^4} - z(m)(2\pi)^D \delta(p)\right] \quad (4)$$

for any smooth $\psi(p)$ such that $\psi(0) = 1$ and $\lim_{p\to\infty} \psi(p) < +\infty$ [4], [6]. Choosing $\psi(p) \equiv 1$ we obtain:

$$z(m) = \int \frac{d^D p}{(2\pi)^D} \frac{1}{m^2+p^2} \equiv \langle T\Phi^2(0) \rangle_0 \neq 0 \quad (5)$$

(note the absence of normal ordering in $\Phi^2(0)$).

Using $\langle RT\Phi^2(0) \rangle_0 = \mu^{D-4} m^2 Z_0 + \langle T\Phi^2(0) \rangle_0$ to express (5) via convergent quantity $\langle RT\Phi^2(0) \rangle_0$ we finally obtain:

$$\Pi(Q,m) = \Pi(Q,0)$$
$$+ \frac{m^2 C_1(Q,\mu) + \langle RT\Phi^2(0) \rangle_0 C_2(Q,\mu)}{Q^2} + o(m^2). \quad (6)$$

Each quantity in (6) is finite *by construction*.

Interaction modifies (6) in three respects:

(i) $C_1$ and $C_2$ become series in coupling constants and $\log(Q/\mu)$;

(ii) $\langle RT\Phi^2(0) \rangle_0$ goes over into $\{\Phi^2(0)\} \equiv \langle RT[\Phi^2(0)S] \rangle_0 \neq 0$, $S$ being the $S$-matrix;

(iii) along with $\{\Phi^2(0)\}$, there appear the vacuum averages $\{\varphi^2(0)\}$ and $\{\varphi(0)\Phi(0)\}$ with their own coefficient functions, due to a mixing mechanism which is analogous to the mixing of operators under UV renormalization.

A generalization of (6) reads:

$$i\int e^{iqx} dx \{A(x)B(0)\} \underset{Q^2\to+\infty}{\cong} \sum_{i=0}^{\infty} \sum_j C_{ij}(Q,\mu) m^{2i} \{O_j(0)\}. \quad (7)$$

$\{O_j(0)\}$ are vacuum averages of local composite MS-renormalized operators [8], inevitably non-vanishing at least while $m \neq 0$. The leading power behavior of $C_{ij}$ at $Q \to \infty$ is determined by power counting and is modified by $\mu$ only via the $\log(Q/\mu)$ contributions. For explicit formulae for calculating $C_{ij}(Q,\mu)$ see [9].

---

[1] The full and general proof will be given elsewhere; it repeats, in essence, the derivation below step by step.

[2] "Correct" means (i) "IR convergent" and (ii) "ensuring the $o(m^2)$ estimate for the neglected terms in (2)".

*3.* Several conclusions can be drawn from the above derivation:

(i) The techniques of dimensional regularization and minimal subtraction [7] are crucial for achieving the neat separation of $Q$ and $\mu$ in (7) (the phenomenon of "perfect factorization" in terms of [6]; see also [10]).

(ii) The quantities $\{O_j\}$ in (7) summarize information on the small-$p$ behavior of the theory. So the above derivation is, in fact, a formalization of heuristic arguments of [1] which led to introduction of non-zero vacuum condensates into deep-Euclidean QCD amplitudes (for a renormalization group analysis of $\{O_j\}$ see below).

(iii) By re-expressing the MS-renormalized operators in (7) in terms of, say, Zimmermann's normal products [11] with zero vacuum averages, all the $\log(Q/\mu)$ terms could only be pumped into the coefficient functions but never got rid of. It should be clearly understood that *the vacuum condensates $\{O_j\}$ are factorization-prescription dependent quantities*, and the virtue of the MS-scheme consists *not* in making them non-zero, but in collecting within $\{O_j\}$ *all* the logs of $m/\mu$.

(iv) It is clear from the derivation that if there are heavy particles in the theory and/or there are other large momenta (in the case of three- etc. -point functions), all the heavy, i.e. $O(q^2)$, parameters will enter only coefficient functions in (7), so that (7) represents a general form of how the world of "light" confined particles affects the "heavy" world. That is to say, *whenever the introduction of vacuum condensates is justified from the phenomenological viewpoint of* [1], *they do appear as an effect of light quark masses already at the perturbative level.*

*4.* Now, the question arises: does the perturbative analysis of the vacuum condensates $\{O_j\}$ make sense?

Consider the set $O_d$ of the aggregates $m^{2i}\{O_j\}$ with a given total dimensionality $d$ in units of mass (they contribute to the same power of $Q^{-2}$ and form a closed set with respect to renormalization). The RG equation for $O_d$ reads:

$$\left( m\frac{\partial}{\partial m} - \beta^*(\alpha_S)\frac{\partial}{\partial \alpha_S} - \left[\frac{d-\gamma(\alpha_S)}{1-\beta_m(\alpha_S)}\right]\right) O_d = 0, \quad (8)$$

where $\gamma$ is the matrix of anomalous dimensions, $\beta^*(\alpha) = \beta(\alpha)(1-\beta_m(\alpha))^{-1}$, and

$$\beta_m(\alpha_S) = \mu\frac{dm}{d\mu}\bigg|_{m_B,\alpha_{SB}} \quad (9)$$

Eq. (8) is similar in form to the well-known evolution equations for moments of structure functions of DIS (for a review see e.g. [12]). The solution to (8) reads:

$$O_d(m) = O_d(m_0)\left(\frac{m}{m_0}\right)^d\left[\frac{\alpha^*(m)}{\alpha^*(m_0)}\right]^{-a_1} \times \left[1 + a_2\alpha^*(m) + \ldots\right], \quad (10)$$

where $\alpha^*(m)$ is governed by $\beta^*$. Assuming that $\alpha^*(m) \approx \left(b\log(m^2/\Lambda^{*2})\right)^{-1}$ with $\Lambda^* \sim \Lambda_{\overline{MS}} \sim 100 \div 500$ MeV, the conclusion is unavoidable that with $O_d(m)$ and light quarks we are in a deep non-perturbative region.

*5.* The most important point perhaps is that even if the $m^d$ factor in (10) prevails in the chiral (i.e. $m_q \to 0$) limit, e.g. if $\langle\bar{q}q\rangle_0$ in the above interpretation vanishes as $m_{u,d} \to 0$, such a possibility is by no means inconsistent with the current-algebra-PCAC phenomenology, as is explained in the appendix E of [13]. As explained there, if $\langle\bar{q}q\rangle_0$ vanishes in the chiral limit then the Gell-Mann–Okubo formula for the $\eta$-meson[3] is lost, while the masses of the quarks $u$ and $d$ are larger than in the standard framework, which gives one a possibility to account for the large deviations from the chiral-invariant picture observed in the pion-nucleon scattering.

However, theoretical estimates of the vacuum condensate values from (10) are hardly feasible at present — one of the reasons is that the value of $\Lambda/\Lambda^*$ is related to the strong-coupling behavior of $\beta_m$, which is unknown.

*6.* To conclude, we have demonstrated that upon resumming large logs of the ratio $m_q/\mu$, where $m_q$ represents light quark masses and $\mu$ is the renormalization [=factorization] parameter, within usual perturbation theory, a collection of "vacuum condensates", i.e. non-zero vacuum averages of composite operators, emerges whenever they are liable to appear according to, and in exactly the same form as prescribed by, the phenomenological rules of [1]. The renormalization group analysis shows that with those condensates we are in a strong-coupling region, so that their behavior at $m_q \to 0$ is hardly predictable at present. However, the fact that the $\log(m_q/\mu)$'s fit into the pattern of vacuum condensates is in itself very nice, for it ensures that those uncalculable terms, being hidden within condensates, cannot modify the formulae of [1], whatever is the interpretation.

*7. Acknowledgements.* The author is grateful to S.G. Gorishny and V.F. Tokarev for discussions. The author is also indebted to K.G. Chetyrkin for promptly turning the author's attention to diagrammatic interpretation of the consistency conditions, Eq. (5), and to J. Gasser (Bern Univ.) for an enlightening discussion of the chiral limit.

---

[3] And nothing else, as was kindly explained to me by J. Gasser.

## Comments (December, 1998)

The subject of the comments below is to explain the systematic approach to studying the structure of power corrections which was demonstrated in this Letter and which remains the only valid and completely meaningful scenario for studying the structure (not necessarily numerical values) of power-suppressed corrections.[4]

### Bibliographic comments

This Letter was in the first burst of publications ([4], [6], [9]) on what eventually evolved into the theory of Asymptotic Operation (hereafter AO; see [19] and refs. therein, and [25] for the more recent ultimate extension to arbitrary non-Euclidean situations). The formulas discovered within the framework of this vast and powerful theory formed the basis of an entire industry of multiloop calculations (NNL moments of DIS structure functions, [23] and many subsequent publications; mass expansions, [30] and refs. therein[5]).

### Remarks on the general method

The derivation presented in sec. 2 follows the pattern of AO explicitly worked out to all orders in [14], [15] (for systematic discussions see [15] and the review [19]). At the time of writing of this Letter, the notion of AO had not yet crystallized out from concrete calculations, instead I emphasized the "extension principle" which is a central but somewhat abstract element in prescriptions for writing out formulas of AO for concrete cases. As to the $R^*$-operation mentioned in the first paragraph of sec. 2, it was actually obtained with an essential use of the method of AO (as explained in [19]) but final formulas could be published without reference to any derivation method (as a calculational trick that just worked), which gave me a possibility to "bootstrap" the publications of the theory of AO — a non-trivial task given a complexity and utter novelty of the theory (it deviated from the then prevailing BPHZ paradigm from the ground up; for more on this see [19]). In the final version of the theory, no reference to the $R^*$-operation is necessary, of course, and the latter remains just an exotic application. In particular, no knowledge of the $R^*$-operation is required to understand the example.

Unfortunately, I had to rely on the method of dimensional regularization (regularization-independent understanding of the method was achieved after this Letter went to press). The use of dimensional regularization somewhat obscured the underlying elementary analytical mechanism.

### The same example without dimensional regularization

A regularization-independent treatment of AO to all orders was accomplished between 1984 [14] and 1990 (see the summary publications [20]). This allows me to explain the example of the Letter without the somewhat confusing use of dimensional regularization which nullifies certain terms that play an important role in the mechanism of factorization; the same mechanism works to all orders [20].

The starting point is now the following 4-dimensional analog of the integral in (1):

$$F = \lim_{\Lambda \to \infty} \left[ \int^\Lambda \frac{d^4 p}{(2\pi)^4} \frac{1}{(p-Q)^2(m^2+p^2)} - z \ln(\Lambda/\mu_{\text{UV}}) \right], \quad 0.1$$

where: $z$ is a finite constant which can be chosen independent of $m$ (this is important; see below); $\Lambda$ is an intermediate upper cutoff; $\mu_{\text{UV}}$ parameterizes the arbitrariness of UV subtraction (the subscript is introduced to explicitly distinguish it from the factorization parameter).

My purpose here is to simply exhibit the factorization mechanism for power correction in the final answer, not explain how such a final answer is arrived at via a deterministic procedure of AO (for that, see [15]). So I am going to simply present the final answer in parallel with the formulas in the main text — an expansion of $F$ including one power term in which the dependence on the large parameter, $Q$, is cleanly separated ("factorized") from the dependence on $m$.

Since the dependence on $\mu_{\text{UV}}$ is logarithmic, it is irrelevant whether one considers $Q$ as large or $m$ as a small parameter. As usual in the theory of AO, it is more convenient to work with small parameters, i.e. expand in $m \to 0$.

Instead of the integral (2), we now write:

$$F_0 \equiv \lim_{\Lambda \to \infty} \left[ \int^\Lambda \frac{d^4 p}{(2\pi)^4} \frac{1}{(p-Q)^2} \times \frac{1}{p^2} - z \ln(\Lambda/\mu_{\text{UV}}) \right]$$
$$+ \int^\infty \frac{d^4 p}{(2\pi)^4} \left\{ \frac{1}{(p-Q)^2} - \frac{\theta(p^2 < \mu^2)}{(0-Q)^2} \right\} \left\{ -\frac{m^2}{p^4} \right\}. \quad 0.2$$

To understand the structure of this expression, compare it with the formal expansion of the $m$-dependent propagator:

$$\frac{1}{m^2 + p^2} = \frac{1}{p^2} - \frac{m^2}{p^4} + \cdots \quad 0.3$$

In 0.2, the UV subtraction is inherited by the integral corresponding to the first term on the r.h.s. of 0.3, $p^{-2}$, whereas the second term, $m^2 p^{-4}$, is accompanied by another subtraction and the cutoff $\theta(p^2 > \mu)$, which play a role similar to that of dimensional regularization and minimal subtraction in the main text. $\mu$ will be seen to be the factorization parameter.

---

[4] Concerning the discussion of the chiral limit in sec. 5 of the main text I only add that, given how nearly impossible it is to do and publish work that goes against a prevailing dogma created by many fine experts, I would not be surprised if the possibility of quark condensates vanishing in the chiral limit (although at a slower than canonical rate; cf. (10)) would turn out to be correct in the end. But that is only a possibility which one can neither prove nor completely exclude.

[5] Unfortunately, the publications by Kühn et al. systematically fail to give proper credit to the original publications in which the mass expansion formulas they used were discovered, namely, [14]–[18]. This is the more deplorable that the resulting misinformation of the research community as regards the true source of the underlying mathematical expertise proves to be detrimental to the progress of theoretical high energy physics because the full potential of the theory of AO is far from being exhausted (cf. [31]).

Eq. (5) is now to be replaced by the following expression:

$$z(m) = \int^\infty \frac{d^4p}{(2\pi)^4} \left[ \frac{1}{m^2+p^2} - \frac{1}{p^2} + \frac{m^2}{p^4}\theta(p^2>\mu^2) \right], \qquad 0.4$$

which is finite both at small and large $p$. One easily finds by a rescaling $p \to mp$ that

$$z(m) = m^2\left[c_1\ln(m/\mu) + c_0\right], \qquad 0.5$$

where both $c_i$ are numerical constants.

A similar power-and-log dependence on $Q^2$ (with an overall $Q^{-2}$) is obtained for the second integral in 0.2 via rescaling $p \to |Q|p$.

Now the "factorized" expansion is as follows:

$$F = F_0 + \int^\infty \frac{d^4p}{(2\pi)^4} \frac{1}{(p-Q)^2} \{z(m)\delta(p)\} + o(m^2)$$

$$= F_0 + z(m) \times \int^\infty \frac{d^4p}{(2\pi)^4} \frac{1}{(p-Q)^2} \{\delta(p)\} + o(m^2). \qquad 0.6$$

(In fact, the remainder is not just $o(m^2)$ [i.e. vanishes faster than $m^2$] but is actually $O(m^4 \ln m)$.)

Correctness of the result 0.6 (i.e. the indicated smallness of the remainder) is easily checked by combining the r.h.s. and l.h.s. and simple algebraic rearrangements of various terms.

The r.h.s. of 0.6 together with 0.4 and the r.h.s. of 0.2 represents a decomposition of the initial integral 0.1 into pieces (without explicit integrations) in such a way as to obtain an expansion in powers and logarithms of the expansion parameter (small $m$ or, equivalently, large $Q$). Prior to discussing the interpretation of various terms in 0.6, let us focus on the pattern of subtractions in the above formulas.

"Lazy cutoffs"

(i) Note how the cutoffs like $\theta(p^2>\mu^2)$ enter the various terms: $\theta(p^2>\mu^2)$ is only introduced precisely into those subtracted terms that correspond to the singularities of the formal expansion, whereas the non-singular term $p^{-2}$ *carries no cutoff*. For instance, the expression for $z(m)$ should be regarded (this interpretation remains valid in all orders of perturbation theory) as obtained by subtracting from the integrand all those and only those asymptotic terms for $p \to \infty$ that are responsible for UV divergence, whereas the cutoff is introduced only into the term corresponding to the logarithmic singularity which necessarily also has a logarithmic singularity at $p=0$, is why the cutoff is needed in the first place. In other words, the finiteness of the corresponding integrals is ensured by subtraction of power-behaved terms rather than a hard cutoff. This mechanism can be conveniently called *lazy cutoff*.

(ii) Recall how factorization is discussed in the language of "factorization theorems" (cf. [21], [22]). There, one discourses (cf. [21]) upon contributions of various integration regions, and the factorization parameter $\mu$ is interpreted as the boundary of various integration regions. However, a precise mechanism of factorization as exhibited in our example[6] is such that the "factorized" integrals collect contributions *from all $p$*.

(iii) So, the interpretation of factorization in terms of "regions" is fundamentally flawed in one trivial respect: information in an integral comes not only from the region but first and foremost from the *integrand*. This flaw manifests itself at the level of non-leading powers:

(iv) Indeed, if one starts with splitting the initial integral into two regions: $|p|<|\mu|$ and $|p|>|\mu|$ (recall that in this example we work with a Euclidean integral), and performs the corresponding expansions ($Q \to \infty$ and $m \to 0$, respectively) then the results would contain various (negative and positive) powers of $\mu$, and keeping track of their eventual cancellations as well as reassembling the correct factorization result 0.6, is hard already in one loop and is a hopeless endeavor in higher orders of PT where the overall mechanism becomes rather delicate; cf. [20].

(v) Plainly put, the mechanism of factorization based on hard cutoffs fails with power-suppressed corrections and — unlike an implicit belief behind the "theory of renormalons" [21] — cannot be a basis for a meaningful theory of such corrections.

Discussion of Eq. 0.6

(i) Eq. 0.6 is in exact correspondence with Eq. (6): $\Pi(Q,0)$ is exactly the first integral in the expression for $F_0$, the r.h.s. of 0.2; $m^2 C_1(Q,\mu)Q^{-2}$ corresponds to the second integral on the r.h.s. of 0.2; $C_2(Q,\mu)Q^{-2}$ is exactly the integral with $\delta$-function on the r.h.s. of 0.6; and the aggregate $\langle RT\Phi^2(0)\rangle_0$ is identified with the $z(m)$, Eq. 0.4.

(ii) The integral with the $\delta$-function exactly corresponds to the much used formulas for coefficient functions of OPE found in a companion Letter [6] and worked out in [9] (cf. the 3-loop calculation [23] that defined the state of the art in DIS OPE calculations).

(iii) The correspondence $\langle RT\Phi^2(0)\rangle_0 \leftrightarrow z(m)$ is a special case of the general results on the connection of diagram-wise UV renormalization (the $R$-operation) and the subtraction of asymptotics at large loop momenta (properly interpreted) directly from non-integrated integrands [24], [20].

(iv) It is a characteristic feature of the method of AO that the integrals one obtains in the "factorization" formulas allows a direct diagrammatic interpretation in terms of some operators — one never needs any matching conditions.

(v) The all-orders generalization of the discussed example as described in the main text after (6) was achieved in [14]–[18] where, moreover, the results were extended to arbitrary models and arbitrary Euclidean asymptotic regimes, as promised in remark (iv) in sec. 3 of the main text.

(vi) An important update to the remark (i) in sec. 3 of the main text is that the dimensional regularization is not crucial for achieving the perfect factorization: the analysis of [20] shows how the required subtractions (with lazy cutoffs) are rather naturally implemented in a regularization-independent manner.

(vii) In cases such as gluon condensates, the masslessness of perturbative gluons prevents non-zero condensates from appearance in lowest orders (due to absence of a mass scale). However, non-zero contributions re-emerge in higher orders (massive quark self-energy insertions).

(viii) A theoretical estimate of the aggregates such as vacuum condensates comprising the large logarithms in power-suppressed corrections requires non-perturbative methods. For Euclidean type objects (e.g. local vacuum condensates), the lattice QCD may provide such estimates. In non-Euclidean situations, no meaningful method exists.

---

[6] Expansions in perfectly factorized form are essentially unique [15], so one cannot talk about different equivalent representations here.

Scenario for non-Euclidean regimes

The scheme used in the main text of the Letter can be summarized as follows:
— perform "perfect factorization" of small parameters such as quark masses;
— identify contributions which give rise to large non-perturbative logarithms;
— treat such PT-uncalculable contributions as phenomenological parameters.

In other words, the small masses etc. are used as a test of stability of perturbation theory in order to identify the structure of potentially large non-perturbative corrections.

The argument that the condensates one thus obtains are not "the" condensates of [1] is beyond the point for two reasons:

1) the large $\ln m$ contributions cannot be phenomenologically distinguished from contributions of, say, instantons;

2) the usual analyses such as [1] *also rely on PT results in regard of the structure of power-suppressed corrections* (the OPE used in [1] was the perturbative OPE somewhat dramatically misinterpreted due to ignoring the non-zero large $\ln m$ contributions which in Zimmermann's version of OPE are hidden in coefficient functions); in this respect our scenario is only a straightforward clarification of the reasoning behind [1].

This scenario is applied to non-Euclidean situations in a straightforward manner (as was mentioned in [26] in connection with the physics of jets). All one needs is a generalization of the method of AO that allows treatment of arbitrary asymptotic regimes and phase space integrals along with loops; such a generalization was found in [25]. One proceeds as follows:

• One starts from complete diagrams with non-zero masses for quarks, …, non-zero $p_\perp$, etc.

• One performs diagram-by-diagram factorization of the small parameters (say, $<1\,\text{GeV}$) using the algorithms of non-Euclidean AO [25]. Note that the method of AO yields "perfect" expansions in which the large logarithms of small parameters are localized in certain integrals. This is in contrast with the conventional methods; e.g. the version of OPE due to Zimmermann had such logarithms hidden in coefficient functions, which phenomenologically is incorrect (see [19]).

• Among various pieces of the resulting factorized expression, one identifies (using scalings like those used to obtain 0.5) the integrals which comprise the large logarithms of small masses. The simple argument $\ln Q/m \to \ln Q/\mu - \ln m/\mu$ shows that if one normalizes $\alpha_S$ at $\mu$ of order $Q$, a typical scale of the process being studied, then the logarithms of masses are automatically large, so neither analysis, nor even knowledge of the corresponding evolution equations is really necessary (the data on power corrections are never precise enough to warrant that).

• One classifies independent "non-perturbative" integrals (i.e. ones that give rise to large logarithms) which would then correspond to independent operator objects (being lower-orders contributions to matrix elements of the latter). Note that interpreting such objects in, say, coordinate representation as local etc. operators serves no useful purpose except, perhaps, as an inessential notational convenience. Anyway, even if one decides that one needs such an operator interpretation, one only needs to *recognize* the emerging integrals (by simply looking at them) as corresponding to some or other operator objects (*not necessarily* local operators) — one does not need to massage them in any way. Their regrouping in a gauge invariant way may be necessary though (e.g. via studying dependence on the gauge parameter).

• Note that for identifying the contributing operators, *matching conditions are never needed*. Note also that if one follows the prescriptions of [25] carefully, one never has to invent a priori prescriptions for renormalization of the resulting non-local operators (cf. [32]) — the pattern of necessary subtractions is dictated by the method of AO unambiguously (apart from inessential finite ambiguities similar to the difference between MS and MS-bar renormalization schemes).

• Upon identification and classifications of independent "non-perturbative" integrals, one simply treats them as phenomenological constants.

• One should replace such integrals with constants as a whole, i.e. without attempting to scale out powers of small masses because such a possibility is a special feature of perturbation theory. That a typical scale in such integrals should be $\Lambda$ rather than, say, quark masses was clear since at least the studies of QCD via Dyson equations in the early 80s [29]. Artificial arguments such as based on Borel resummation of Borel-non-resummable QCD PT series [21] cannot add anything new here as a matter of principle.

• If one has trouble identifying independent integrals, the best solution is to consider the next order of PT (one *does not need* to do the integrals explicitly, only write out the prescriptions of AO [25] explicitly, which requires only patience) where radiative corrections would allow one to distinguish different operator objects.

• To facilitate investigation of contributions involving gluon operators (which to lowest order may be zero due to perturbative masslessness of the gluon), one may force an early appearance of non-zero gluon operators by introducing a gluon mass (concerning gauge invariance of such a procedure consult [27]). This is just a trick because quark self-energies would give non-zero contributions from higher orders anyway.

• The examples I studied show that the breakdown of dimensional regularization observed in [28] in a semi-artificial example, is quite typical in the case of power corrections to, say, jet observables, so one must exercise care with intermediate regularizations.

• The success of the QCD sum rules method [1] where one dealt with essentially Euclidean regimes and was able to parameterize many observables with a few phenomenological constants, is not likely to be repeated in non-Euclidean situations where one encounters a wider variety of non-perturbative operator objects in addition to local operators and has to deal with unknown phenomenological *functions* of kinematic arguments, sometimes in infinite number.

• Moreover, whereas vacuum condensates (that are the only type of non-perturbative coefficients in Euclidean situations) can be estimated from, say, the lattice QCD, for the non-Euclidean situations no useful and meaningful method exists.

• Finally, a critical examination of renormalons-based results demonstrates that their concrete content is always traced back to (rather imperfect) analysis of diagrams, whereas the renormalon phraseology effectively reduces to the unoriginal proposition that the corresponding aggregates are non-perturbative without providing any clue as to how one could compute them.